# Calibrating Function Points Using Neuro-Fuzzy Technique

Vivian Xia

IT Department
HSBC Bank Canada
Vancouver, BC, Canada

vivian_xia@hsbc.ca

Danny Ho

NFA Estimation Inc.
London, Ontario, Canada

danny@nfa-estimation.com

Luiz F. Capretz

Department of Electrical and
Computer Engineering
University of Western Ontario
London, Ontario, Canada

lcapretz@eng.uwo.ca

**Abstract**

The concepts of calibrating Function Points are discussed, whose aims are to fit specific software application, to reflect software industry trend, and to improve cost estimation. Neuro-Fuzzy is a technique which incorporates the learning ability from neural network and the ability to capture human knowledge from fuzzy logic. The empirical validation using ISBSG data repository Release 8 shows a 22% improvement in software effort estimation after calibration using Neuro-Fuzzy technique.

## 1. Introduction

Function Points (FP) is an ideal software size metric to estimate cost since it can be obtained in the early development phase, such as requirement, measures the software functional size from user's view, and is programming language independent [1], [2]. To achieve more accurate estimation, we suggest the concepts of calibration which are detailed in the next section.

Neural network technique is based on the principle of learning from historical data. The neural network is trained with a series of inputs and desired outputs from the training data set [3]. Once the training is complete, new inputs are presented to the neural network to predict the corresponding outputs. Fuzzy logic is a technique to make rational decisions in an environment of uncertainty and imprecision. It is rich in representing human linguistic ability with the terms such as fuzzy set, fuzzy rules [4]–[6]. Once the concept of fuzzy logic is incorporated into the neural network, the result is a Neuro-Fuzzy system that combines the advantages of both techniques [7], [8]. This technique is found appropriate to calibrate Function Points as proved by the validation results.

## 2. Concepts of Calibration

### 2.1 Calibrate to fit specific application

In Function Points counting, each component, such as Internal Logical File (ILF), is classified to a complexity determined by its associated files numbers, such as Data Element Types (DET), Record Element Types (RET) [2] as listed in Table 1. Such complexity classification is easy to operate, but it may not fully reflect the software complexity under the specific software application. For example, Table 2 shows a software project with three ILFs: A, B and C. According to the complexity matrix, A and B are classified as having the same complexity and are assigned the same weight value of 10. However, A has 30 more DET than B and is certainly more complex, but they are now assigned the same complexity. Also, B is classified as average and assigned a weight of 10 while C is classified as low and assigned a weight of 7. B has only one more DET than C and the same number of RET as C, but B has been assigned three more weight units than C. There is no smooth transition boundary between two classifications. Processing the number of Function Points component associated files such as DET, RET using fuzzy logic can produce an exact complexity degree.

**Table 1: ILF Complexity Matrix.**

| ILF | DET | | |
|---|---|---|---|
| RET | 1-19 | 20-50 | 51+ |
| 1 | Low | Low | Avg |
| 2-5 | Low | Avg | High |
| 6+ | Avg | High | High |

**Table 2: Observations of FP Complexity Classification.**

| | ILF A | ILF B | ILF C |
|---|---|---|---|
| DET | 50 | 20 | 19 |
| RET | 3 | 3 | 3 |
| Complexity | Average | Average | Low |
| Weight Value | 10 | 10 | 7 |

2.2 <u>Calibrate to reflect industry trend</u>

The weight values of Unadjusted Function Points (Table 3) are said to reflect the functional size of software [1]. They were determined by Albrecht in 1979 based on the study of 22 IBM Data Processing projects. Since 1979, software development has been growing steadily and is not limited to one organization or one type of software. Thus, there is a need to calibrate these weight values to reflect the current software industry trend. The International Software Benchmarking Standard Group (ISBSG) maintains a large empirical project data repository. ISBSG data repository Release 8 contains 2,027 projects, which cover a broad range of project types with 75% of the projects being less than 5 years old. Learning UFP weight values from ISBSG Release 8 using neural network calibrates Function Points to reflect the current software industry trend.

**Table 3: UFP Weight Values.**

| Component | Low | Average | High |
|---|---|---|---|
| External Inputs | 3 | 4 | 6 |
| External Outputs | 4 | 5 | 7 |
| External Inquiries | 3 | 4 | 6 |
| Internal Logical Files | 7 | 10 | 15 |
| External Interface Files | 5 | 7 | 10 |

2.3 Calibrate to improve cost estimation

The significant relationship between the software size and cost has been recognized for a long time. In the classical view of cost estimation process (Figure 1), the outputs of *effort* and *duration* are estimated from software *size* as the primary input and a number of *cost factors* as the secondary inputs. There are mainly two types of software size metrics: Source Lines of Code (SLOC) and Function Points. SLOC is a natural artifact that measures software physical size, but it is usually not available until the coding phase and difficult to have the same definition across different programming languages. Function Points is an ideal software size metric to estimate cost since it can be obtained in the early development phase, such as requirement, measures the software functional size, and is programming language independent [2]. Calibrating Function Points incorporates the historical information and gives a more accurate view of software size. Hence more accurate cost estimation comes with a better software size metric.

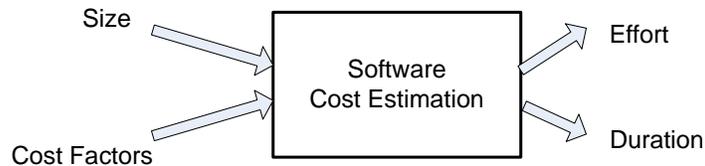

**Figure 1: Classical View of Cost Estimation Process.**

## 3. Neuro-Fuzzy Calibration Approach

We propose an approach to calibrate Function Points using Neuro-Fuzzy technique. The model overview and two parts of the model: fuzzy logic part and neural network part are described here. The empirical validation is provided in the next section.

3.1 Overview

The block diagram shown in Figure 2 gives an overview of our approach. The project data provided by ISBSG [9] is imported to extract an estimation equation and to train the neural network. An estimation equation is extracted from the data set by statistical regression analysis. Fuzzy logic is used to calibrate Function Points complexity degree to fit specific application. Neural network calibrates UFP weight values to reflect the current software industry trend by learning from ISBSG data. The validation results show that the calibrated Function Points have better estimation ability than that of the original.

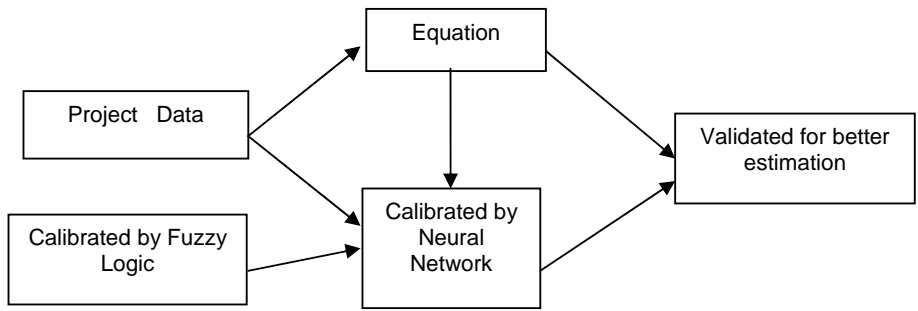

**Figure 2: Block Diagram of Neuro-Fuzzy Approach.**

### 3.2 Fuzzy Logic Part

The fuzzy logic part calibrates the Function Points complexity degree to fit the specific application. A fuzzy logic system (shown in Figure 3) is constructed based on the fuzzy set, fuzzy rules and fuzzy inference. The input fuzzy sets are to fuzzify the component associated file numbers and the output fuzzy set are to fuzzify the complexity classification. The fuzzy rules are defined in accordance with the original complexity weight matrices. The fuzzy inference process using the Mamdani approach [10] is applied based on the fuzzy sets and fuzzy rules.

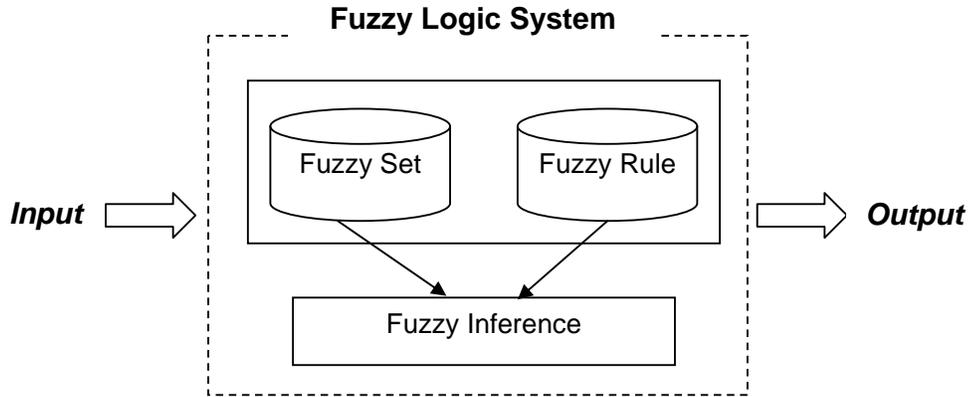

**Figure 3: Fuzzy Logic System.**

### 3.3 Neural Network Part

The neural network part is aiming at calibrating Function Points to reflect the current software industry trend. By learning from ISBSG data repository, this part is believed to achieve the calibration goal. In order to reach a reasonable conclusion, the raw ISBSG data set is filtered by several criteria recommended by ISBSG [11]. A subset of 184 projects is obtained of which the quality rating is A or B, the counting method is IFPUG which excludes other counting methods such as COSMIC FFP [12] and Mark II [13], the effort resource is recorded at level one (development team), the development type is new development or re-development, the 15 Unadjusted Function Point breakdowns and 14 General System Characteristics rating values are available.

The neural network is constructed to receive 15 UFP breakdowns as inputs to give the work effort as the desired output. A back-propagation learning algorithm [3] is conducted in order to minimize the prediction difference between the estimated and actual efforts. An effort estimation equation is extracted based on the data subset using statistical regression analysis. The equation in the form of ***Effort = A· UFP $^B$*** is achieved with the help of the statistical software SPSS v12 [14].

### 4. Validation Results

Five experiments were conducted to validate our Neuro-Fuzzy approach. For each experiment, the original data set (184 projects) was randomly separated into 100 training data points and 84 test data points. The outliers are the abnormal project data points with large noise that may distort the training result. Thus, we used the training data set excluding the outliers for calibration, but used the rest of the data points for validation [15]. The average calibrated UFP weight values obtained from five experiments are listed in Table 4, and the original weight values are given as comparison.

**Table 4: Calibrated UFP Weight Values.**

| Component | Low | | Average | | High | |
|---|---|---|---|---|---|---|
| | Original | Calibrated | Original | Calibrated | Original | Calibrated |
| External Inputs | 3 | 0.9 | 4 | 2.2 | 6 | 4.7 |
| External Outputs | 4 | 3.3 | 5 | 4.6 | 7 | 6.2 |
| External Inquiries | 3 | 1.8 | 4 | 2.9 | 6 | 5.4 |
| Internal Logical Files | 7 | 5.4 | 10 | 9.8 | 15 | 14.9 |
| External Interface Files | 5 | 4.6 | 7 | 6.9 | 10 | 10 |

The validation results of the five experiments are assessed by Mean Magnitude Relative Error (MMRE) for estimation accuracy. MMRE is defined as: for $n$ projects, $MMRE = \frac{1}{n}\sum_{i=1}^{n}\left(|Estimated_i - Actual_i|/Actual_i\right)$. The validation results of the five experiments are listed in Table 5 where "Improvement %" is the MMRE improvement in percentage for each experiment. Based on the MMRE assessment results, an average of 22% cost estimation improvement has been achieved with the Neuro-Fuzzy calibration approach. The MMRE after calibration is over 100% which is still relatively large and is due to the absence of well-defined cost drivers like COCOMO [16] [17] factors. Unfortunately ISBSG Release 8 does not have data on cost drivers.

**Table 5: MMRE Validation Result.**

| | Exp.1 | Exp.2 | Exp.3 | Exp.4 | Exp.5 |
|---|---|---|---|---|---|
| **MMRE Original** | 1.38 | 1.58 | 1.57 | 1.39 | 1.42 |
| **MMRE Calibrated** | 1.10 | 1.28 | 1.17 | 1.03 | 1.11 |
| **Improvement %** | 20% | 19% | 25% | 26% | 22% |
| **Average Improvement %** | 22% | | | | |

The validation results of the five experiments are also assessed by Prediction at level $p$ (PRED) criteria, i.e., $PRED(p) = k/N$, where $N$ is the total number of projects, $k$ is the number of projects with absolute relative error of $p$. Four PRED criteria are assessed here, namely Pred 25, Pred 50, Pred 75 and Pred 100. Table 6 lists the Pred assessment result where the overall performance is improved.

**Table 6: PRED Validation Results.**

| | Average Original | Average Calibrated | Average Improvement |
|---|---|---|---|
| **Pred 25** | 13% | 12% | -1% |
| **Pred 50** | 23% | 27% | 4% |
| **Pred 75** | 40% | 46% | 6% |
| **Pred 100** | 60% | 67% | 7% |

## 5. Conclusion

The Neuro-Fuzzy approach to calibrate Function Points is validated with the empirical data repository (ISBSG Release 8). The experimental validation results show a 22% improvement in software cost estimation and demonstrate that Function Points need calibration and can be calibrated. The fuzzy logic part of the model calibrates the Function Points complexity degree to fit the specific application context. The neural network part of the model calibrates the UFP weight values to reflect the current software industry trend. The combined neuro-fuzzy technique calibrates Function Points for better cost estimation.